\documentclass{emulateapj}

\begin{document}

\title{Spatially Resolved \textit{Spitzer} Spectroscopy of the
        Starburst Nucleus in NGC 5253} 

\author{P. Beir\~ao\altaffilmark{1},
        B. R. Brandl\altaffilmark{1}, 
        D. Devost\altaffilmark{2},
        J. D. Smith\altaffilmark{3}, 
        L. Hao\altaffilmark{2} \&
        J. R. Houck\altaffilmark{2}} 
\altaffiltext{1}{Sterrewacht Leiden, P. O. Box 9513, 2300 RA Leiden, The Netherlands} 
\altaffiltext{2}{Astronomy Department, Cornell University, Ithaca, 
                 NY 14853-6801} 
\altaffiltext{3}{Steward Observatory, University of Arizona, Tucson, 
                 AZ 85721}

\begin{abstract}
We present new {\sl Spitzer Space Telescope} data on the nearby,
low-metallicity starburst galaxy NGC~5253, from the Infrared Array
Camera {\sl IRAC} and the Infrared Spectrograph {\sl IRS}\footnote{The {\sl IRS} was a collaborative venture between Cornell University and Ball Aerospace Corporation funded by NASA through the Jet
Propulsion Laboratory and the Ames Research Center.}.  The mid-IR
luminosity profile of NGC~5253 is clearly dominated by an unresolved
cluster near the center, which outshines the rest of the galaxy at
longer wavelengths.  We find that the [Ne\,III]/[Ne\,II] ratio
decreases from $\sim8.5$ at the center to $\sim2.5$ at a distance of
$\sim250$~pc.  The [S\,IV]/[S\,III] follows the [Ne\,III]/[Ne\,II]
ratio remarkably well, being about $4-5$ times lower at all distances.
Our spectra reveal for the first time PAH emission feature at 11.3$\mu$m and its
equivalent width increases significantly with distance from the
center.  The good anti-correlation between the PAH strength and the
product between hardness and luminosity of the UV radiation field
suggests photo-destruction of the PAH molecules in the central region.
The high-excitation [OIV]25.91$\mu$m line was detected at
$0.42\times10^{-20}$W cm$^{-2}$. 
Our results demonstrate the importance of spatially resolved mid-IR spectroscopy.

\end{abstract}

\keywords{ISM: lines and bands - Galaxies: individual (NGC 5253) - Galaxies: ISM - Galaxies: starburst - Infrared: galaxies}

\newpage

\section{Introduction}

NGC~5253 is a nearby, low-metallicity dwarf galaxy with a recent
starburst, which is responsible for its infrared luminosity of
$L_{IR}\sim1.8\times10^9L_{\odot}$ \citep{Beck96}.  Distance estimates
vary from $3.3\pm0.3$~Mpc \citep{Gibson00} to $4.0\pm0.3$~Mpc
\citep{Thim03}; here we will assume 4.0~Mpc, corresponding to
19.4~pc/$\arcsec$.  With its low metallicity of only about $1/6
Z_{\odot}$ \citep{Kobul99} NGC~5253 is an excellent target to study
starbursts in a low-metallicity environment.  The spectral signatures
of Wolf-Rayet (WR) stars suggest a very recent starburst
(\citet{Beck96,Schaerer97}).  \citet{Cresci05} detected 115 star
clusters using optical and near-infrared VLT images at an age range of
3-19~Myr.  \citet{Turner98} found a compact radio source representing
a hidden super star cluster (SSC) in one of the earliest phases of SSC
formation ever observed. Its ionizing flux corresponds to several
thousand O7\,V star equivalents within the central $2''$
\citep{Crowther99,Turner04} and an infrared luminosity of
$L_{IR}=7.8\times10^8L_{\odot}$ \citep{Crowther99}.  Near-infrared
observations with Hubble Space Telescope revealed the presence of a
double star cluster in the nuclear region, separated by $6-8$~pc
\citep{Alonso04}. There are indications that an interaction with M81 might have provoked the starburst \citep{Kobul95}.

NGC~5253 has also been studied in the mid-IR with the \textit{Infrared
Space Observatory (ISO)} by several authors, e.g., \citet{Crowther99},
\citet{Thornley00} and \citet{Verma03}.  With {\sl Spitzer}'s
increased sensitivity \citep{Werner04} and the smaller slit apertures,
the {\sl IRS} can continue where {\sl ISO} left off.  In this letter
we report on the spatial variations of the physical conditions in the
central region of NGC~5253, based on {\sl IRAC} images and {\sl IRS} \citep{Houck04a}
spectral maps.

\begin{figure}
\epsscale{.80}
\plotone{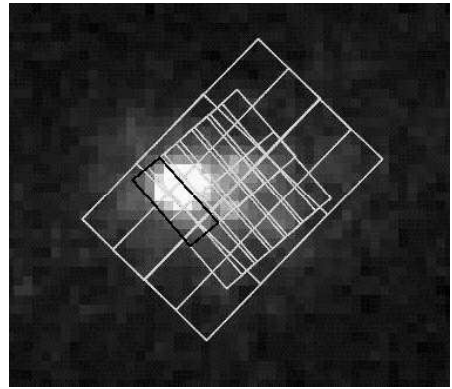}
\caption{Overlay of the SH and LH slits on the {\sl IRAC} 3.6$\mu$m
         image of NGC~5253. The image is $1.2'\times1.0'$. Highlighted is the SH slit at the most central position.}
\end{figure}

\section{Observations and Data Reduction}

The images were obtained on 2005 January 31 using IRAC
\citep{Fazio04} at all four bands ($3.6, 4.8, 5.8, 8.0\mu$m). The
observations consist of 12 slightly dithered pointings of $3\times
12$s exposures each. The data were pipeline processed by the Spitzer
Science Center. Longward of $5\mu$m the IRAC images show only one compact cluster
and no structure of the host galaxy.  To assess what fraction of the total
luminosity of the central region is provided by the central cluster we
compare the flux within the central 44~pc (2 pixel) to the total flux
within a radius of 330 pc for each channel.  The results in
Table 1 show the increasing dominance of the starburst
nucleus in luminosity with wavelength.  We subtracted the {\sl IRAC}
instrumental PSFs from the nucleus for all four channels, and the
residuals suggest that the central cluster remains unresolved in the
{\sl IRAC} images, which is consistent with its very compact
size \citep{Turner04}.

\begin{deluxetable}{cccc}
\tablewidth{0pt}
\tablecaption{IRAC Flux Table}
\tablehead{\colhead{$\lambda$} & \colhead{central flux} & 
           \colhead{total flux} & \colhead{$F_{44pc}/F_{600pc}$} \\ 
           \colhead{($\mu$m)} & \colhead{(MJy/sr)} & \colhead{(MJy/sr)} & 
           \colhead{}} 
\startdata
3.6 & 2616$\pm$12 & 6578$\pm$86 & 0.39 \\
4.5 & 5069$\pm$16 & 7846$\pm$97 & 0.65 \\
5.8 & 11640$\pm$29 & 17187$\pm$261 & 0.68 \\
8.0 & 26869$\pm$33 & 37846$\pm$99 & 0.71 \\
\enddata
\end{deluxetable}

The mid-IR spectra were obtained on 2004 July 14, using high
resolution modules ($R\approx 600$) of the {\sl Infrared Spectrograph
(IRS)} in spectral mapping mode.  In the SH
(short-high) mode, the map consists of 12 different pointings,
overlapping by half a slit width and about one third slit length,
covering an area of $18.0\arcsec\times 23.6\arcsec$.    In LH (long-high) mode the map
consists of only 6 different pointings, covering an area of $22.2\arcsec\times33.4\arcsec$. Fig. 1 shows the SH
and LH slit positions overlaid on the {\sl IRAC} image of the central
region.  Both maps are slightly off-center.  Additional ``sky''
measurements, 6~arcmin from the nucleus were also taken.
The basic processing of the data was performed with version~11.0 of
the automated {\sl IRS} pipeline at the Spitzer Science Center. The background was subtracted using the sky images.  
The spectra were extracted in "full-slit" mode from pre-flat-fielded
files using the {\sl IRS} data reduction and
analysis package SMART, version 5.5 \citep{Higdon04}.  The extracted
spectra were flux calibrated with an empirically derived RSRF 
(relative spectral response function) of
$\alpha$ Lac. The spectral overlaps between orders were manually clipped,
according to the local S/N.  Finally, the SH spectra were scaled up by
16\% to match the LH continuum fluxes at 19~$\mu$m. This discrepancy in the 
fluxes is due to the difference between the SH and LH slit sizes.


\section{Discussion}

The complete $10-38\mu$m SH+LH spectrum at the most central position
of our map is shown in Fig. 2.  With a
smooth continuum with no significant absorption features, and turnover
in the slope around $20\mu$m, it is dominated by the strong emission
lines of [SIV]$10.5\mu$m, [NeII]$12.8\mu$m, [NeIII]$15.5\mu$m,
[SIII]$18.7\mu$m, [SIII]$33.5\mu$m and [SiII]$34.8\mu$m. Also detected are the signature of
polycyclic aromatic hydrocarbons (PAH) at $11.3\mu$m and Hu$\alpha$ at $12.37\mu$m up to a distance of $\sim120$ pc.
Fig. 3 shows eight representative SH spectra at
decreasing radial distances, calculated from the
central cluster to the center of each slit position. The remaining
four spectra of our map are redundant and not shown to save space.
The most prominent spectral features are labeled. The line fluxes for
each radial position are listed in Table 2.

\begin{figure}
\plotone{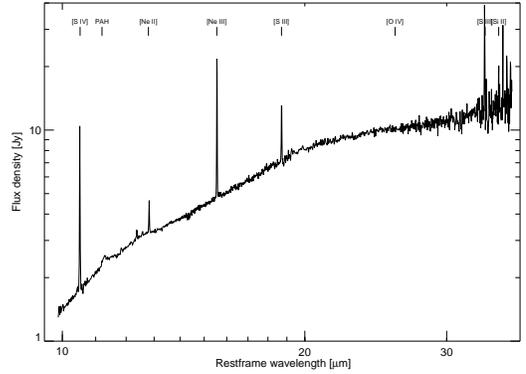}
\caption{Complete {\sl IRS} SH+LH spectrum of the cluster region of
         NGC~5253, which correspond to the highlighted slits in Fig. 1. Beyond 35$\mu$m the spectrum dominated by noise and detector artifacts.}
\end{figure}

\begin{figure*}
\plotone{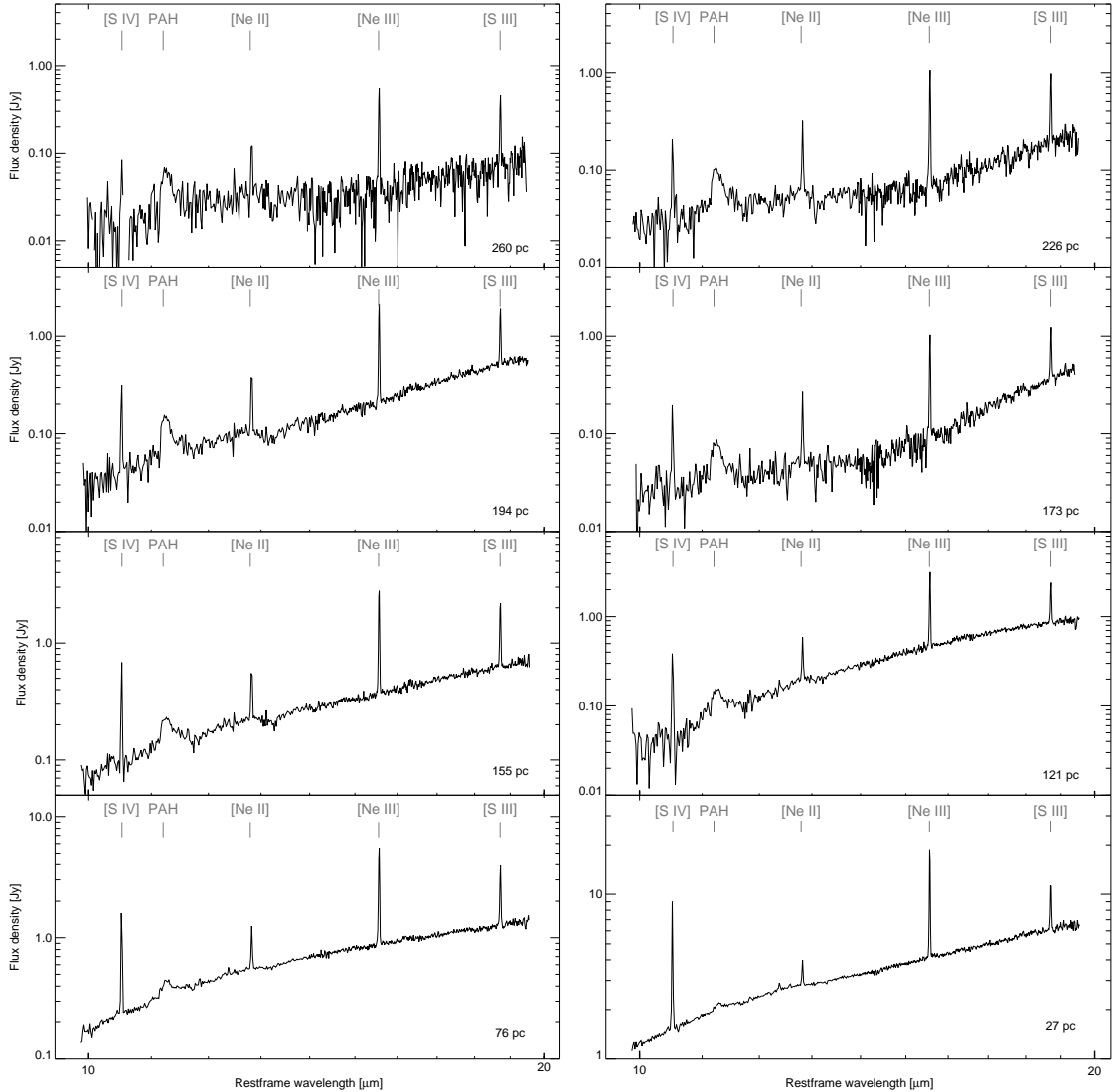}
\caption{IRS SH spectra of NGC~5253 sorted by decreasing distance from
         the central cluster (from left to right and from top to
         bottom). The distances are inserted in the lower right of
         each plot.}
\end{figure*}


\subsection{Gradients in the Radiation Field}

With ionization potentials of 21.56~eV and 40.95~eV for Ne and
Ne$^{+}$, respectively, the [Ne\,III]/[Ne\,II] ratio is a good measure
of the hardness of the radiation field, and traces the OB stars. \citet{Crowther99} measured a relatively low [Ne\,III]/[Ne\,II] ratio
of $3.5-4.0$.  However, the larger {\sl ISO-SWS} slit apertures of
$14''\times 20''$ and $14''\times 27''$, centered on the nucleus, may
have also picked up significant line flux at lower excitation from the
surrounding galactic population.  With the smaller {\sl IRS} slit
aperture of $4.7''\times 11.3''$ we can probe the influence of
spatial resolution on the measured spectral diagnostics.  The upper
plot of Fig. 4 shows the [Ne\,III]/[Ne\,II] and
[S\,IV]/[S\,III] ratios as functions of the distance to the central
cluster.  Both ratios decrease by a factor of four over 250 parsecs,
indicating a significant softening of the UV radiation field with
distance from the cluster core, and trace each other remarkably well, with the Ne ratio being
about $4-5$ times higher at all distances. Our high [Ne\,III]/[Ne\,II] ratio at the central cluster is comparable
to those observed in nearby H\,II regions like 30~Doradus and
low-metallicity dwarf galaxies like II~Zw~40 \citep{Thornley00}.
\citet{Rigby04} modeled Ne line ratios for star clusters at low
metallicity $Z=0.2Z_{\odot}$ and showed that a peak ratio of 7.0 is
consistent with an upper mass cutoff of $100 M_{\odot}$, at an age of
$3-5$~Myr.


\subsection{Dependence of PAH Strength on the Radiation Field}

It has often been asked if PAHs in low-metallicity starbursts appear
to be weaker because of low abundance or because they get destroyed by
the generally harder radiation fields in these environments (e.g.,
\citet{Wu05,Engel05,Halloran05,Houck04b,Madden00}). Our high S/N spectra
clearly reveal, for the first time, the presence of the $11.3\mu$m PAH
feature in all SH positions on NGC~5253, which shows that PAHs can be
present in a low metallicity environment.  The center plot of
Fig. 4 shows a steady increase of the PAH equivalent
width (EW) with distance. We assume constant metallicity
throughout the galaxy, as no known dwarf has
steep metallicity gradients \citep{Kobul99a}.
In Fig. 4 we have investigated the correlation between
the measured PAH strength and the ``strength'' of the UV radiation
field, defined by the product of the hardness and the intensity of the radiation field, [Ne\,III]/[Ne\,II]$\times$([Ne\,III]+[Ne\,II]). The bottom plot shows the product between the UV field and the
PAH strengths as a function of distance.  This product stays almost
constant out to a radial distance of 200 pc, meaning that the strength
of the UV field and the strength of the PAH emission is strongly
anti-correlated.  The good anti-correlation over such a large distance
(encompassing numerous H\,II regions) suggests that the
photo-destruction of PAHs could be the dominant mechanism here. We note that PAH emission models \citep{Li2001,Bakes01} show that photoionization of hydrogen in PAHs can also cause a decrease in the relative PAH strength above $10\mu$m. However, we consider this effect unlikely since no ionization effect has been seen in other starburst galaxies \citep{Brandl06}. For comparison, a low metallicity system like SBS 0335-052, where no PAHs have been detected, has a Ne ratio of 4.9 \citep{Houck04b} and some of the low metallicity systems with weak PAHs studied by \citet{Halloran05} have a Ne ratio of $\sim3-4$. As seen in Fig. 4, these ratios correspond to a radial distance inside the PAH destruction zone in NGC 5253.


\subsection{[O\,IV] Line Emission}

With an excitation potential of
54.9~eV, the [O\,IV]25.89$\mu$m line fills the wide energetic gap of mid-IR fine-structure lines
between lines that can originate from massive stars and lines that likely require an AGN. It has been attributed to various mechanisms, including very hot stars \citep{Schaerer99,Morris04} and energetic shocks for low-excitation starbursts \citep{Lutz98}. Our high S/N spectra reveal a faint [O\,IV] line at two slit positions outside the central cluster,
with fluxes of about $0.42\times10^{-20}$W cm$^{-2}$ at a S/N of 7.1. This is higher than the limits given by \citet{Crowther99} ($0.10\times10^{-20}$W), but below
the upper limit of \citet{Verma03} ($0.9\times10^{-20}$W cm$^{-2}$), and in reasonable agreement with the
flux measured by \citet{Lutz98} ($0.65\times10^{-20}$W cm$^{-2}$). Using the
STARBURST99 code (Leitherer et al. 1999) assuming an instantaneous burst
of star formation with a Salpeter \citep{Salpeter55} IMF at 1/5 solar
metallicity, the observed OIV emission can be produced by roughly 125 WR
stars (WC+WN), consistent with the wide range of O7\,V star
equivalents \citep{Crowther99,Turner04} within the central $1-2''$. A detailed discussion is
given in \citet{Martin05}. However, as the [OIV] emission is only observed in two LH slit positions off the nucleus, 
it is not obvious that the [OIV] line is predominantly photo-excited by the central WR stars. Other excitation mechanisms, such as shocks, need to be considered.

\begin{figure}
\epsscale{.80}
\plotone{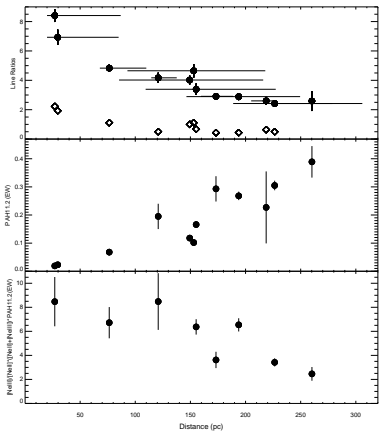}
\caption{\textit{Upper:} Variation of the [Ne\,III]/[Ne\,II] (full
         circles) and [S\,IV]/[S\,III] (diamonds) line ratios with
         distance to the galactic center. The error bars along the x
         axis represent the {\sl IRS} slit length, which translates
         into a range in radial distance; they are the same for
         [S\,IV]/[S\,III] ratios. The error bars along the y axis
         represent flux uncertainties. \textit{Center:} PAH $11.2
         \mu$m equivalent widths (EWs) as a function of distance to
         the nucleus. The error bars represent EW measurement
         uncertainties. \textit{Bottom:} Dependence of PAH $11.2\mu$m
         EW $\times$ ([Ne\,II]+[Ne\,III]) $\times$ [Ne\,III]/[Ne\,II] with
         distance. For details see text.  The bars indicate systematic
         errors.}
\end{figure}

\begin{deluxetable*}{cccccccc}
\tabletypesize{\scriptsize}
\tablecaption{Fine-Structure Lines and Ratios\label{tablines}}
\tablehead{ 
\colhead{distance} & \colhead{[SIV]} & \colhead{Hu$\alpha$} & \colhead{[NeII]} &
\colhead{[NeIII]} & \colhead{[SIII]} & \colhead{[NeIII]/[NeII]} &
\colhead{[SIV]/[SIII]} \\ 
\colhead{(pc)} & \colhead{$(10^{-20} Wcm^{-2})$} & \colhead{$(10^{-20}
Wcm^{-2})$} & \colhead{$(10^{-20} Wcm^{-2})$} & \colhead{$(10^{-20}
Wcm^{-2})$} & \colhead{$(10^{-20} Wcm^{-2})$} \\ 
\colhead{} & \colhead{$\lambda_{obs}=10.52\mu$m} &
\colhead{$\lambda_{obs}=12.37\mu$m} &
\colhead{$\lambda_{obs}=12.83\mu$m} & 
\colhead{$\lambda_{obs}=15.58\mu$m} &
\colhead{$\lambda_{obs}=18.74\mu$m} & \colhead{} & \colhead{} \\
\colhead{} & \colhead{EP=34.8 eV} & \colhead{} & \colhead{EP=21.6 eV} &
\colhead{EP=41.0 eV} & \colhead{EP=23.3 eV} & \colhead{} & \colhead{}
}
\startdata

27 & 37.50 $\pm$ 0.21 & 2.11 $\pm$ 0.53 & 5.63 $\pm$ 0.14 & 47.40 $\pm$ 1.21 & 16.79 $\pm$ 0.58 & 8.41 & 2.23 \\

30 & 33.22 $\pm$ 0.84 & 1.57 $\pm$ 0.28 & 5.90 $\pm$ 0.16 & 40.87 $\pm$ 1.90 & 17.24 $\pm$ 0.52 & 6.93 & 1.93 \\

76 & 8.80 $\pm$ 0.21 & 0.43 $\pm$ 0.02 & 3.51 $\pm$ 0.08 & 16.94 $\pm$ 0.50 & 7.91 $\pm$ 0.24 & 4.83 & 1.11 \\

121 & 2.25 $\pm$ 0.32 & 0.32 $\pm$ 0.10 & 2.02 $\pm$ 0.09 & 8.41 $\pm$ 0.31 & 4.54 $\pm$ 0.16 & 4.17 & 0.50 \\

135 & 7.01 $\pm$ 0.45 & \nodata & 2.62 $\pm$ 0.17 & 12.18 $\pm$ 0.33 & 6.30 $\pm$ 0.18 & 4.65 & 1.11 \\
 
149 & 6.03 $\pm$ 0.51 & \nodata & 2.75 $\pm$ 0.13 & 11.06 $\pm$ 0.25 & 6.00 $\pm$ 0.12 & 4.02 & 1.01 \\
  
155 & 3.37 $\pm$ 0.24 & \nodata & 2.58 $\pm$ 0.23 & 8.74 $\pm$ 0.14 & 4.98 $\pm$ 0.09 & 3.39 & 0.68 \\

173 & 1.01 $\pm$ 0.08 & \nodata & 1.09 $\pm$ 0.02 & 3.16 $\pm$ 0.11 & 2.38 $\pm$ 0.08 & 2.91 & 0.43 \\
 
194 & 1.77 $\pm$ 0.04 & \nodata & 2.17 $\pm$ 0.14 & 6.27 $\pm$ 0.09 & 4.03 $\pm$ 0.09 & 2.89 & 0.44 \\

219 & 0.71 $\pm$ 0.01 & \nodata & 0.60 $\pm$ 0.04 & 1.55 $\pm$ 0.06 & 1.10 $\pm$ 0.07 & 2.60 & 0.64 \\
 
226 & 1.18 $\pm$ 0.18 & \nodata & 1.36 $\pm$ 0.08 & 3.28 $\pm$ 0.11 & 2.35 $\pm$ 0.09 & 2.42 & 0.50 \\
 
260 & \nodata & \nodata & 0.68 $\pm$ 0.12 & 1.76 $\pm$ 0.07 & 1.12 $\pm$ 0.05 & 2.59 & \nodata 
\enddata

\end{deluxetable*}

\acknowledgments

This work is based on observations made with the {\em Spitzer} Space
Telescope, which is operated by the Jet Propulsion Laboratory,
California Institute of Technology under NASA contract 1407. Support
for this work was provided by NASA through Contract Number 1257184
issued by JPL/Caltech.

\end{document}